\documentclass[12pt]{article}
  \usepackage{amsthm,amsfonts}
  \usepackage{amsmath}
\newcommand{\bea}   {\begin{eqnarray}}
\newcommand{\eea}   {\end{eqnarray}}
\usepackage{subeqnarray}
\usepackage[brazilian]{babel}
\usepackage[latin1]{inputenc}
\usepackage{url}
\begin{document}
\renewcommand{\thefootnote}{\fnsymbol{footnote}}
\date{}

\baselineskip=17pt
\begin{center}
{\bf Ensaio sobre a unidimensionalidade do tempo}
\end{center}

\begin{center}
{\it F. Caruso}\footnote{Pesquisador titular do Centro Brasileiro de Pesquisas F\'{\i}sicas e Professor Colaborador no Programa de P\'{o}s-Gradua\c{c}\~{a}o em Hist\'{o}ria das Ci\^{e}ncias e das T\'{e}cnicas e Epistemologia (HCTE) da Universidade Federal do Rio de Janeiro (UFRJ).}
\end{center}

\vspace*{1.cm}

	``\textit{A qualidade do espa\c{c}o e do tempo, por exemplo, que o primeiro tem tr\^{e}s dimens\~{o}es e o segundo, somente uma, s\~{a}o princ\'{\i}pios que} (...)'' [1]. Assim quis o acaso que chegasse at\'{e} n\'{o}s a conclus\~{a}o \`{a} qual o Kant maduro chegou ao revisitar um tema que lhe interessou desde a juventude: o problema da dimensionalidade. Esse fragmento, publicado em seu \textit{Opus Postumum}, est\'{a} irreversivelmente interrompido em um ponto crucial.

	Em seu \textit{The Natural Philosophy of Time}, Whitrow dedica menos de cinco p\'{a}ginas \`{a} possibilidade do tempo ser pluridimensional [2]. Sua breve apresenta\c{c}\~{a}o do tema gira em torno de dois pontos, essencialmente: a quest\~{a}o do conhecimento pr\'{e}vio e a da revers\~{a}o de perspectiva. A primeira ligada ao sonho, por exemplo, enquanto a segunda refere-se \`{a} possibilidade de se ver um simples desenho de um cubo ora em uma perspectiva, ora em outra. A conclus\~{a}o \`{a} qual chega \'{e} de que n\~{a}o h\'{a} nenhuma raz\~{a}o convincente de que o tempo tenha mais do que uma dimens\~{a}o. Al\'{e}m disso, afirma que
\begin{quotation}
\noindent ``\textit{\'{e} dif\'{\i}cil acreditar que nosso sistema da F\'{\i}sica, baseado no conceito de uma vari\'{a}vel tempo unidimensional, possa ter tanto sucesso quanto tem se de fato habit\'{a}ssemos um mundo no qual o tempo tivesse duas ou mais dimens\~{o}es}''.
\end{quotation}

	O objetivo desse ensaio \'{e} apresentar ao leitor como essa discuss\~{a}o pode ser travada no \^{a}mbito da F\'{\i}sica. Para tanto, come\c{c}aremos revendo alguns aspectos do problema da dimensionalidade do espa\c{c}o.

	Desde 1983, o autor tem se dedicado ao estudo de aspectos hist\'{o}ricos e filos\'{o}ficos do \textit{espa\c{c}o} e, em particular, vem procurando compreender as origens de sua dimensionalidade, investigando como esse atributo topol\'{o}gico relaciona-se \`{a} estrutura de v\'{a}rias leis f\'{\i}sicas [3-17]. Dentre tantas an\'{a}lises, feitas com diversos colaboradores, podemos citar alguns exemplos de fen\^{o}menos f\'{\i}sicos que dependem do n\'{u}mero de dimens\~{o}es do espa\c{c}o no qual eles t\^{e}m lugar: a atra\c{c}\~{a}o regida pelo potencial $1/r$ (na gravita\c{c}\~{a}o cl\'{a}ssica ou na eletrost\'{a}tica) [14-17], a difra\c{c}\~{a}o de n\^{e}utrons [3], o efeito Casimir [4,9], o espectro estelar [7] e a radia\c{c}\~{a}o de fundo [11]. Em todos estes trabalhos -- como em todos os outros que tratam do problema da dimensionalidade do espa\c{c}o -- o tempo \'{e} pressuposto ser \textit{unidimensional}. Assim, mesmo sabendo-se que um particular evento f\'{\i}sico, na realidade, ocorre no \textit{espa\c{c}o-tempo}, se est\'{a}, na verdade, discutindo e impondo limites ou v\'{\i}nculos sobre a dimensionalidade apenas do espa\c{c}o, como \'{e} feito, por exemplo, em [18]. Essa \'{e} uma estrat\'{e}gia limitada de abordar o problema.

	Na verdade, esta limita\c{c}\~{a}o n\~{a}o deve causar espanto, uma vez que discutir a dimensionalidade do espa\c{c}o ou do tempo a partir de singularidades que as leis f\'{\i}sicas possam apresentar em rela\c{c}\~{a}o a um particular n\'{u}mero de dimens\~{o}es esbarra sempre no fato de que tais leis s\~{a}o sempre determinadas te\'{o}rica ou empiricamente sem qualquer tipo de questionamento \textit{a priori} da dimensionalidade seja do espa\c{c}o, seja do tempo. \'{E} como se estas fossem um dado de fato da Natureza, uma verdade inquestion\'{a}vel. No caso do tempo, essa esp\'{e}cie de preconceito parece estar ainda mais arraigada na comunidade cient\'{\i}fica, como sugere, por exemplo, a refer\^{e}ncia feita anteriormente a Whithrow.

	A experi\^{e}ncia sens\'{\i}vel de ordena\c{c}\~{a}o temporal e a unidimensionalidade do tempo parecem t\~{a}o imbricadas que, de fato, a literatura sobre o problema da dimensionalidade temporal \'{e} muito reduzida comparada \`{a}quela que trata do problema an\'{a}logo para o espa\c{c}o. Podemos, inclusive, questionar se o pr\'{o}prio conceito de causalidade que herdamos n\~{a}o depende de se ter tais rela\c{c}\~{o}es como verdadeiras. Sendo assim, em que fatos experimentais ou em que outros conceitos basilares poderia se basear quem estiver interessado em justificar que o tempo \'{e} unidimensional ou mesmo provar que pode ser multidimensional?

Talvez possamos afirmar que a origem do car\'{a}ter unidimensional do tempo remonta ao abandono do tempo c\'{\i}clico difundido no pensamento grego cl\'{a}ssico, a partir do surgimento do cristianismo. Mais especificamente, com Santo Agostinho, que, apesar de sua consagrada resposta \`{a} quest\~{a}o do que \'{e} o tempo -- ``Se ningu\'{e}m mo perguntar eu sei; se o quiser explicar a quem me fizer a pergunta, j\'{a} n\~{a}o sei.'' [19] -- foi quem apresentou a primeira teoria filos\'{o}fica do tempo, baseada numa origem (a crucifica\c{c}\~{a}o de Cristo) e na convic\c{c}\~{a}o de que o tempo \'{e} a medida, pela consci\^{e}ncia humana, do movimento ``retil\'{\i}neo'' da hist\'{o}ria, irrevers\'{\i}vel e que n\~{a}o se repete [20].

	Antes de esbo\c{c}armos algumas contribui\c{c}\~{o}es no sentido de rever brevemente o que foi feito at\'{e} hoje buscando compreender a dimensionalidade do tempo, gostar\'{\i}amos de destacar o fato de que uma parte significativa dos argumentos que concernem \`{a} dimensionalidade do espa\c{c}o depende da exist\^{e}ncia de um espa\c{c}o m\'{e}trico [3]. Tal fato nos remete \`{a} no\c{c}\~{a}o de dist\^{a}ncia em uma variedade $n$ dimensional, a qual se baseia tradicionalmente na forma diferencial homog\^{e}nea quadr\'{a}tica
$$ \mbox{d} s^2 = g_{\mu\nu} \mbox{d}x^\mu \mbox{d}x^\nu$$

\noindent na qual os \'{\i}ndices $\mu$ e $\nu$ assumem os valores $0, 1, 2, \cdots (n-1)$ . Essa f\'{o}rmula, em \'{u}ltima an\'{a}lise, \'{e} uma escolha arbitr\'{a}ria, pois, de fato, n\~{a}o h\'{a} argumentos l\'{o}gicos que excluam \textit{a priori} outras formas do tipo $\mbox{d} s^4, \mbox{d} s^6, \mbox{d} s^8...$.  Neste ponto \'{e} importante lembrar que, em 1920, Paul Ehrenfest apresentou a conjectura de que o expoente 2 da forma quadr\'{a}tica na equa\c{c}\~{a}o anterior para o elemento de linha poderia estar relacionado com a dimensionalidade do espa\c{c}o [14] mas, at\'{e} onde sabemos, tal conjectura ainda n\~{a}o foi demonstrada. Algumas consequ\^{e}ncias, como uma poss\'{\i}vel rela\c{c}\~{a}o dessa conjectura com o teorema de Fermat, foram discutidas em [3]. Al\'{e}m disto, o fato de muitas das equa\c{c}\~{o}es fundamentais da F\'{\i}sica envolverem derivadas espaciais de segunda ordem (a equa\c{c}\~{a}o de Newton, a equa\c{c}\~{a}o de onda de d'Alembert, a equa\c{c}\~{a}o de Schr\"{o}dinger \textit{etc.}) pode tamb\'{e}m estar relacionado \`{a} tridimensionalidade do espa\c{c}o. Voltarei a este ponto mais adiante.
	
Na realidade, foi o desenvolvimento das Geometrias n\~{a}o-Euclidianas no s\'{e}culo XIX [21] que permitiu as primeiras especula\c{c}\~{o}es a cerca de uma quarta dimens\~{a}o e o que ela seria, bem antes da Teoria da Relatividade, dentre as quais posso citar a de Hinton [22]. Outros aspectos do problema da realidade ou n\~{a}o de uma quarta dimens\~{a}o, ligados \`{a} percep\c{c}\~{a}o e \`{a} filosofia foram tratados por Whitrow [2]. De qualquer forma, do ponto de vista da F\'{\i}sica, \'{e} a m\'{e}trica da geometria de Minkowski que pode ser facilmente generalizada para um n\'{u}mero qualquer de dimens\~{o}es espaciais e temporais. Para $\mu$  e $\nu$  variando de 0 a 3 ($n = 4$), tem-se a m\'{e}trica

 $$ g^{\mu \nu} = \left(
                    \begin{array}{cccc}
                      + & 0 & 0 & 0 \\
                      0 & - & 0 & 0 \\
                      0 & 0 & - & 0 \\
                      0 & 0 & 0 & - \\
                    \end{array}
                  \right)$$
\vspace*{0.3cm}

Para um n\'{u}mero qualquer $n$ de dimens\~{o}es do espa\c{c}o-tempo, a nova matriz $g^{\mu\nu}$  ter\'{a} dimens\~{o}es $n \times n$. Em seu famoso livro \textit{The Mathematical Theory of Relativity}, Arthur Stanley Eddington pondera que tal escolha (um sinal $+$ e tr\^{e}s $-$) ``\textit{particulariza o mundo de um modo que dificilmente poder\'{\i}amos ter predito a partir de primeiros princ\'{\i}pios}'' [23]. Por que o espa\c{c}o-tempo tem uma e n\~{a}o outra assinatura? Lembra, ent\~{a}o, o astrof\'{\i}sico ingl\^{e}s, sem citar a refer\^{e}ncia, que Hermann Weyl expressa este car\'{a}ter ``especial'' afirmando que o espa\c{c}o tem   dimens\~{o}es [24]. Entretanto, uma leitura atenta de seus trabalhos mostra que esse seria o n\'{u}mero total de dimens\~{o}es espa\c{c}o-temporais que assegura a invari\^{a}ncia de escala do Eletromagnetismo Cl\'{a}ssico de Maxwell, mas com a dimensionalidade do tempo pr\'{e}-fixada em~1.

De volta ao livro de Eddington, ele examina outra quest\~{a}o interessante: se o universo pode mudar sua geometria. Em particular, se pergunta se em alguma regi\~{a}o remota do espa\c{c}o ou do tempo se poderia ter uma m\'{e}trica do tipo

$$ g^{\mu \nu} = \left(
                    \begin{array}{cccc}
                      - & 0 & 0 & 0 \\
                      0 & - & 0 & 0 \\
                      0 & 0 & - & 0 \\
                      0 & 0 & 0 & - \\
                    \end{array}
                  \right)$$

Sua resposta \'{e} negativa e o argumento \'{e} que, se tal regi\~{a}o existe, ela deve estar separada por uma superf\'{\i}cie da regi\~{a}o em que a assinatura da m\'{e}trica \'{e} $(+,-,-,-)$, de tal forma que, para um lado da superf\'{\i}cie de separa\c{c}\~{a}o, tem-se

$$ \mbox{d} s^2 = c_1^2\mbox{d} t^2 - \mbox{d} x^2 - \mbox{d} y^2 - \mbox{d} z^2$$

enquanto que, do outro lado,

 $$ \mbox{d} s^2 = - c_2^2\mbox{d} t^2 - \mbox{d} x^2 - \mbox{d} y^2 - \mbox{d} z^2$$

A transi\c{c}\~{a}o, nesse caso, s\'{o} poderia ocorrer atrav\'{e}s de uma superf\'{\i}cie na qual

 $$ \mbox{d} s^2 = 0 \mbox{d} t^2 - \mbox{d} x^2 - \mbox{d} y^2 - \mbox{d} z^2$$

Portanto, a velocidade (fundamental) c da luz seria nula, do que resulta que

\begin{quotation}
\noindent ``\textit{Nada poderia se mover na superf\'{\i}cie de separa\c{c}\~{a}o entre as duas regi\~{o}es e nenhuma influ\^{e}ncia pode passar de um lado para outro. A suposta ulterior regi\~{a}o n\~{a}o est\'{a} em qualquer rela\c{c}\~{a}o espa\c{c}o-temporal com nosso universo -- o que \'{e} um modo de certa forma pedante de dizer que ela n\~{a}o existe}'' [23].
\end{quotation}

Hoje se sabe que nas teorias cl\'{a}ssicas da Gravita\c{c}\~{a}o n\~{a}o pode haver altera\c{c}\~{o}es locais na topologia do espa\c{c}o-tempo sem que se considerem flutua\c{c}\~{o}es qu\^{a}nticas [25]. O caso de um mundo hipot\'{e}tico de dimens\~{o}es $2+2$ \'{e} tamb\'{e}m brevemente discutido pelo autor. A possibilidade de um universo no qual o tempo possa ser bidimensional \'{e} tratada ainda em outro livro de Eddington publicado postumamente [26].
	
Em 1970, Dorling desenvolveu um argumento essencialmente cinem\'{a}tico e mostrou que
\begin{quotation}
\noindent ``\textit{a propriedade extrema [m\'{a}xima] de geod\'{e}sicas do tipo tempo em um espa\c{c}o-tempo ordin\'{a}rio \'{e} uma condi\c{c}\~{a}o necess\'{a}ria para a exist\^{e}ncia de part\'{\i}culas est\'{a}veis. Esta propriedade de m\'{a}ximo falharia se o tempo fosse multidimensional}'' [27].
\end{quotation}

Segundo esse autor, para um tempo multidimensional, o pr\'{o}ton e o el\'{e}tron n\~{a}o seriam est\'{a}veis. Nem mesmo o f\'{o}ton! Al\'{e}m disso, prop\~{o}e que as obje\c{c}\~{o}es que se apresentam para velocidades maiores do que a da luz e para um tempo multidimensional podem estar relacionadas. Abre-se, assim, uma possibilidade de estudo dos \textit{tachyons} em universos com um n\'{u}mero maior de coordenadas do tipo tempo. Isto porque a \'{u}nica diferen\c{c}a essencial entre tempo e espa\c{c}o (e entre as correspondentes geod\'{e}sicas do tipo tempo ou do tipo espa\c{c}o) em uma geometria de Minkowski \'{e} a diferen\c{c}a na respectiva dimensionalidade.

	\'{E} tamb\'{e}m digno de nota o trabalho de Mirman [28], no qual ele defende a tese de que a assinatura do espa\c{c}o-tempo parece estar relacionada ao processo de medida e, se houver mais de uma dimens\~{a}o do tipo tempo, as dimens\~{o}es extras n\~{a}o seriam observ\'{a}veis. A quest\~{a}o da medida nos parece ser um ponto central em toda essa discuss\~{a}o sobre o n\'{u}mero de dimens\~{o}es, seja do espa\c{c}o, seja do tempo.

Qualquer processo de medida depende n\~{a}o s\'{o} da defini\c{c}\~{a}o de um observador como tamb\'{e}m de algumas leis f\'{\i}sicas. Frequentemente, o que \'{e} feito para se discutir o problema da dimensionalidade do espa\c{c}o \'{e} generalizar a forma funcional de uma equa\c{c}\~{a}o diferencial que descreve uma lei da F\'{\i}sica em um espa\c{c}o $R^3$ (que -- \'{e} sempre preciso lembrar -- foi estabelecida sem qualquer tipo de questionamento sobre a tridimensionalidade do espa\c{c}o) para um espa\c{c}o $R^n$, mas mantendo-se a ordem da equa\c{c}\~{a}o diferencial. Assim, quando se discute a estabilidade planet\'{a}ria baseada na gravita\c{c}\~{a}o newtoniana em espa\c{c}os de dimens\~{o}es arbitr\'{a}rias o que se faz \'{e} generalizar a equa\c{c}\~{a}o de Poisson do seguinte modo:
$$\nabla^2_{(3)} \phi = \frac{\partial^2}{\partial x_1^2} + \frac{\partial^2}{\partial x_2^2} + \frac{\partial^2}{\partial x_3^2} = 4\pi \rho \ \ \Rightarrow \ \ \nabla^2_{(n)} \phi = \frac{\partial^2}{\partial x_1^2} + \frac{\partial^2}{\partial x_2^2} + \cdots + \frac{\partial^2}{\partial x_n^2} = 4 \pi \rho $$

A partir da\'{\i} acha-se a solu\c{c}\~{a}o geral de equa\c{c}\~{a}o generalizada e, admitindo-se por hip\'{o}tese (no fundo justificada apenas por argumentos de natureza antr\'{o}pica) que ela descreva igualmente bem o mesmo fen\^{o}meno f\'{\i}sico do caso $n=3$, discute-se a estabilidade mec\^{a}nica desta nova solu\c{c}\~{a}o. As limita\c{c}\~{o}es epistemol\'{o}gicas deste m\'{e}todo foram amplamente discutidas em [3].
	
Nesse ponto gostar\'{\i}amos de recordar alguns trabalhos do caro e saudoso amigo Juan Jos\'{e} Giambiagi, que, com Guido Bollini desenvolveu, em 1972, o famoso m\'{e}todo de regulariza\c{c}\~{a}o dimensional [29-30], admitindo que a dimensionalidade do espa\c{c}o-tempo seja um n\'{u}mero real dado por $\nu = 3+1 -\epsilon$. \textit{Bocha} (como era conhecido entre os amigos) trabalhou com diferentes colaboradores sobre o problema da dimensionalidade do espa\c{c}o e do tempo de uma forma muito aberta, sem qualquer tipo de preconceito [31-41]. Nesses trabalhos seminais de 1972, Bollini \& Giambiagi mostraram, pela primeira vez (at\'{e} onde sabemos), que uma pequena flutua\c{c}\~{a}o imposta \textit{ad hoc} \`{a} dimensionalidade do espa\c{c}o-tempo est\'{a} na base de um m\'{e}todo capaz de controlar diverg\^{e}ncias que surgem no c\'{a}lculo de certas quantidades f\'{\i}sicas em teorias de campo de calibre. Portanto, mostraram que a din\^{a}mica em uma teoria de campos tamb\'{e}m pode depender crucialmente do n\'{u}mero de dimens\~{o}es do espa\c{c}o-tempo.
	
Nos artigos seguintes, em linhas muito gerais, Giambiagi e colaboradores d\~{a}o particular \^{e}nfase ao estudo da equa\c{c}\~{a}o de onda de d'Alembert generalizada e sua rela\c{c}\~{a}o com o princ\'{\i}pio de Huygens. O que \'{e} importante para o escopo desse ensaio \'{e} chamar aten\c{c}\~{a}o para o fato de que eles o fazem de uma maneira bem mais geral do que a generaliza\c{c}\~{a}o da equa\c{c}\~{a}o de Poisson anteriormente mencionada, com sofistica\c{c}\~{o}es crescentes a cada artigo, permitindo inclusive novas pot\^{e}ncias para o operador d'alembertiano, $\Box$. O fato de que as propriedades das equa\c{c}\~{o}es de onda dependem fortemente das dimens\~{o}es espaciais n\~{a}o \'{e} novo e j\'{a} havia sido notado por Ehrenfest [14], Henri Poincar\'{e} [42] e Jacques Hadamard [43]. O argumento de que mundos espacialmente tridimensionais parecem ter uma combina\c{c}\~{a}o \'{u}nica e muito peculiar de propriedades que garantam o processamento e propaga\c{c}\~{a}o de sinais via fen\^{o}menos eletromagn\'{e}ticos pode ser encontrado em [44], mas n\~{a}o \'{e} demais frisar que ele \'{e} constru\'{\i}do apoiado na unidimensionalidade do tempo.
	
Motivado por novos desenvolvimentos na Gravita\c{c}\~{a}o e em Teorias Supersim\'{e}\-tri\-cas, Giambiagi busca se libertar ``deste preconceito'' e vai estudar, em v\'{a}rios dos artigos j\'{a} citados aqui, do ponto de vista da F\'{\i}sica Matem\'{a}tica, solu\c{c}\~{o}es para diferentes dimens\~{o}es de equa\c{c}\~{o}es envolvendo os operadores $\Box$, $\Box^{1/2}$, $\Box^2$, $\Box^3$, $\Box^\alpha$, para uma coordenada temporal, inicialmente [37], e depois em um espa\c{c}o-tempo com $(p+q)$ dimens\~{o}es [39, 41], em que
$$\Box = \frac{\partial^2}{\partial t_1^2} + \frac{\partial^2}{\partial t_2^2} + \cdots + \frac{\partial^2}{\partial t_q^2} - \frac{\partial^2}{\partial x_1^2} - \frac{\partial^2}{\partial x_2^2} - \cdots - \frac{\partial^2}{\partial x_p^2}$$

E as solu\c{c}\~{o}es analisadas s\~{a}o aquelas que dependem somente das vari\'{a}veis
$$ t = \sqrt{t_1^2 + t_2^2 + \cdots + t_q^2} \qquad \mbox{e} \qquad r = \sqrt{x_1^2 + x_2^2 + \cdots + x_p^2}$$

\'{E} f\'{a}cil perceber que a natureza epistemol\'{o}gica de um eventual v\'{\i}nculo que se venha a obter destes resultados, mostrando que apenas a pot\^{e}ncia $\alpha = 1$  do operador $\Box$  e um espa\c{c}o-tempo quadridimensional garantiriam a propaga\c{c}\~{a}o de ondas eletromagn\'{e}ticas sem problemas de perda de informa\c{c}\~{a}o e sem reverbera\c{c}\~{o}es [44], seria muito diferente do resultado j\'{a} conhecido. Outras contribui\c{c}\~{o}es mais recentes neste campo da F\'{\i}sica Matem\'{a}tica podem ser encontradas em [45, 46].

Todos esses trabalhos oferecem uma gama substancial de resultados que merecem ser analisados de um ponto de vista epistemol\'{o}gico e n\~{a}o apenas do ponto de vista formal, segundo o qual a multiplicidade de dimens\~{o}es espaciais \'{e} apenas uma possibilidade matem\'{a}tica a ser explorada e investigada. Talvez esse tipo de investiga\c{c}\~{a}o sobre alguns destes resultados possa lan\c{c}ar uma luz sobre o quanto nossa percep\c{c}\~{a}o e a ado\c{c}\~{a}o formal de um tempo unidimensional, por um lado, e as formas funcionais das leis f\'{\i}sicas, por outro, est\~{a}o imbricadas. Ou talvez, como disse Weinstein, ``[teorias com m\'{u}ltiplas dimens\~{o}es espaciais] servem para alargar nossas mentes no sentido do que pode ser fisicamente poss\'{\i}vel''.

\vspace*{1.5cm}

\noindent \textbf{Refer\^{e}ncias bibliogr\'{a}ficas}

\vspace*{0.6cm}

[ 1] E. Kant, \textit{Opus Postumum -- passage des principes m\'{e}taphysiques de la science de la nature \`{a} la physique}, translation, presentation and notes by F. Marty, Paris, Press Univ. de France, 1986, p.~131.

[ 2] G.J. Whitrow, \textit{The Natural Philosophy of Time}. Oxford: University Press, second edition (1980).

[ 3] F. Caruso \& R. Moreira, ``On the physical problem of spatial dimensions: an alternative procedure to stability arguments'', \textit{Fundamenta Scientiae} \textbf{8}, p.~73-91 (1987).

[ 4] F. Caruso, N.P. Neto, B.F. Svaiter \& N.F. Svaiter, ``Attractive or repulsive nature of Casimir force in $D$-dimensional Mikowski spacetime'', \textit{Physical Review D} \textbf{43},
n.~4, p.~1300-1306 (1991).

[ 5]  F. Caruso \& R. Moreira, ``Causa efficiens versus causa formalis: origens da discuss\~{a}o moderna sobre a dimensionalidade do espa\c{c}o'', \textit{Scientia (Unisinos)} \textbf{4}, n.~2,
p.~43-64 (1994).

[ 6] F. Caruso \& R. Moreira, ``Notas sobre o problema da dimensionalidade do espa\c{c}o e da extens\~{a}o no primeiro texto do jovem Kant'', \textit{Scientia (Unisinos)} \textbf{7}, n.~2, p.~13-22 (1996).

[ 7] F. Caruso \& R. Moreira, ``Space dimensionality: what can we learn from stellar spectra and from the M\"{o}ssbauer effect'', in: R.B. Scorzelli, I. Souza Azevedo \& E. Baggio Saitovitch (Eds.), \textit{Essays on Interdisciplinary Topics in Natural Sciences Memorabilia: Jacques A. Danon}, Gif-sur-Yvette/Singapore: \'{E}ditions Fronti\`{e}res, p.~73-84 (1997).

[ 8] F. Caruso \& R. Moreira, ``Sull'influenza di Cartesio, Leibniz e Newton nel primo approccio di Kant al problema dello spazio e della sua dimensionalit\`{a}'', \textit{Epistemologia (Genova, Italia)} \textbf{XXI}, n.~2, p.~211-224 (1998).

[ 9] F. Caruso, R. De Paola \& N.F. Svaiter, ``Zero point energy of massless scalar field in the presence of soft and semihard boundary in $D$ dimensions'', \textit{International Journal of Modern Physics A} \textbf{14}, n.~13, p.~2077-2089 (1999).

\newpage
[10] F. Caruso, ``A note on space dimensionality constraints relied on anthropic arguments: Methane structure and the origin of life'', \textit{In} M.S.D. Cattani; L.C.B. Crispino; M.O.C. Gomes \& A.F.S. Santoro (Eds.) \textit{Trends in Physics: Festschrift in homage to Prof. Jos\'{e} Maria Filardo Bassalo}, S\~{a}o Paulo: Livraria da F\'{\i}sica, p.~95-106 (2009).

[11] F. Caruso \& V. Oguri, ``The Cosmic Microwave Background Spectrum and a Determination o f Fractal Space Dimensionality'', \textit{Astrophysical Journal} \textbf{694}, p.~151-153 (2009).

[12] F. Caruso \& R. Moreira, ``On Kant's first insight into the problem of space dimensionality and its physical foundations'', \url{arXiv:0907.3531v1}, submetido a \textit{Kant Studien} (2010).

[13] I. Kant, in J Handyside (ed.), \textit{Kant's inaugural dissertation and the early writings on space}, Chicago: Open Court, 1929, reimpresso por Hyperion Press (1979).

[14] P. Ehrenfest, ``In what way does it become manifest in the fundamental laws of physics that space has three-dimensions?'', \textit{Koninklijke Nederlandsche Akademie van Wetenschappen Proceedings} \textbf{20}, n.~1, p.~200-209 (1918), reimpresso em M.J. Klein (ed.), \textit{Paul Ehrenfest -- Collected Scientific Papers}, Amsterdam: North-Holland Publ. Co. (1959), 
p.~400-409. Uma c\'{o}pia completa digitalizada est\'{a} dispon\'{\i}vel em \url{http://adsabs.harvard.edu/abs/1918KNAB...20..200E} (acessado em 7 de maio de 2014). Veja tamb\'{e}m, do mesmo autor, ``Welche Rolle spielt die Dreidimensionalit\"{a}t des Raumes in den Grundgesetzen der Physik?'', \textit{Annalen der Physik} \textbf{61}, p.~440-446 (1920).

[15] F.R. Tangherlini, ``Schwarzschild field in n dimensions and the dimensionality of space problem'', \textit{Nuovo Cimento} \textbf{27}, p.~636-651 (1963).

[16] F. Caruso, J. Martins \& V. Oguri, ``On the existence of hydrogen atoms in higher dimensional Euclidean spaces'', \textit{Physics Letters A} \textbf{377} p.~694-698 (2013).

[17] F. Caruso, J. Martins, V. Oguri \& L. Perlingeiro, ``The relativistic hydrogen atom and the dimensionality of space'', em fase de conclus\~{a}o.

[18] B. M\"{u}ller \& A. Sch\"{a}fer, ``Improved bounds on the dimension of space-time'', \textit{Physical Review Letters} \textbf{56}, n.~12, p.~1215-1218 (1986).

[19] Santo Agostinho, \textit{Confiss\~{o}es}, Livro XI.

[20] \textit{Ibid}.

[21] M. Jammer, \textit{Conceitos de Espa\c{c}o: A hist\'{o}ria das teorias do espa\c{c}o na f\'{\i}sica}. Rio de Janeiro, Contraponto e Ed. PUC Rio (2010).

[22] C.H. Hinton, \textit{What is the fourth dimension?}, London: Allen und Unwin (1887). Veja tamb\'{e}m \textit{Speculations on the Fourth Dimension: Selected Writings of Charles H. Hinton}, New York: Dover (1980).

\newpage
[23] A.S. Eddington, \textit{The Mathematical Theory of Relativity}. Cambridge: University Press (1923), p.~25.

[24] H. Weyl, ``Gravitation und Elektrizit\"{a}t'', \textit{Sitzungsberichte der K\"{o}niglich Preu{\ss}ischen Akademie der Wissenschaften zu Berlin}, p.~465-480 (1918). ``Eine neue Erweiterung der Relativit\"{a}tstheorie'', \textit{Annalen der Physik} \textbf{59}, p.~101-133 (1919). Veja tamb\'{e}m seu \textit{Space, Time, Matter}. New York: Dover (1952), p.~282-825.

[25] Veja, por exemplo, J. Martin, N. Pinto-Neto \& I. Dami\~{a}o Soares, ``Green functions for topology change'', \textit{Journal of High Energy Physics} \textbf{3}, p.~060 (2005) e refer\^{e}ncias l\'{a} citadas.

[26] A.S. Eddington, \textit{Fundamental Theory}. Cambridge: University Press (1946), p. 126.

[27] J. Darling, ``The dimensionality of time'', \textit{American Journal of Physics} \textbf{38}, p.~539-540 (1970).

[28] R. Mirman, ``Comments on the dimensionality of time'', \textit{Foundations of Physics} \textbf{3}, p.~321-333 (1973).

[29] C.G. Bollini \& J.J. Giambiagi, ``Dimensional Regularization: The Number of Dimensions as a Regularizing Parameter'', \textit{Nuovo Cimento B} \textbf{12}, p.~20-26 (1972).

[30] C.G. Bollini \& J.J. Giambiagi, ``Lowest order `divergent' graphs in $\nu$-dimen\-sional space'', \textit{Physics Letters B} \textbf{40}, n.~5, p.~566-568 (1972).

[31] C.G. Bollini \& J.J. Giambiagi, ``Supersymmetric Klein-Gordon equation in $d$-dimensions'', \textit{Physical Review D} \textbf{32}, n.~12, p.~3316-3318 (1985).

[32] C.G. Bollini \& J.J. Giambiagi, ``Lagrangian Procedures for Higher order field equation'', \textit{Revista Brasileira de F\'{\i}sica} \textbf{17}, n.~1, p.~14-30 (1987).

[33] C.G. Bollini \& J.J. Giambiagi, ``Higher order equations of Motion'', \textit{Revista Mexicana de F\'{\i}sica} \textbf{36}, n.~1, p.~23-29 (1990).

[34] C.G. Bollini \& J.J. Giambiagi, ``Huyghens' Principle in $(2n+1)$ Dimensions for Nonlocal Pseudodifferential Operator of the Type $\Box^\alpha$'', \textit{Nuovo Cimento A} \textbf{104}, n.~12, p.~1841-1844 (1991).

[35] C.G. Bollini, J.J. Giambiagi \& O. Obreg\'{o}n, ``Are some physical theories related with a specific number of dimensions?'', in A. Feinstein \& J. Ib\'{a}\~{n}ez (Eds.), \textit{Recent Developments in Gravitation} (Proceedings of Spanish Conference on Gravitation), Singapore: World Scientific (1992), p. 103.

[36] C.G. Bollini \& J.J. Giambiagi, ``Criteria to Fix the Dimensionality Corresponding to Some Higher Derivative Lagrangians'', \textit{Modern Physics Letters A} \textbf{7}, n.~7, 
p.~593-599 (1992).

[37] C.G. Bollini \& J.J. Giambiagi, ``Arbitrary Powers of d'Alembertians and the Huygens' Principle'', \textit{Journal of Mathematical Physics} \textbf{34}, n.~2, p.~610-621 (1993).

\newpage

[38] J.J. Giambiagi, ``Relations Among Solutions for Wave and Klein-Gordon Equations for Different Dimensions'', \textit{Nuovo Cimento B} \textbf{109}, n.~6, p.~635-644 (1994).

[39] W. Bietenholz \& J.J. Giambiagi, ``Solutions of the Spherically Symmetric Wave Equation in $p+q$  dimensions'', \textit{Journal of Mathematical Physics} \textbf{36}, n.~1, p.~383-397 (1995).

[40] C.G. Bollini, J.J. Giambiagi, J. Benitez \& O. Obreg\'{o}n, ``Which is the Dimension of Space if Huygens' Principle and Newtonian Potential are Simultaneously Satisfied?'', \textit{Revista Mexicana de F\'{\i}sica} \textbf{39}, suplemento n.~1, p.~S1-S6 (1993).

[41] J.J. Giambiagi, ``Wave Equations with multiple times: Classical and Quantum Solutions'', \textit{preprint} CBPF-NF-055 (1995).

[42] H. Poincar\'{e}, \textit{Derni\`{e}res Pens\'{e}es}. Paris: Flammarion (1917).

[43] J. Hadamard, \textit{Lectures on Cauchy's problem in linear partial differential equations}. New Haven: Yale University Press (1923).

[44] J.D. Barrow, ``Dimensionality'', \textit{Philosophical Transactions of the Royal Society of London A} \textbf{310}, p.~337-346 (1983).

[45] W. Craig \& S. Weinstein, ``On determinism and well-posedness in multiple time dimensions'', \url{arXiv:0812.0210v3 (2009)}. \textit{Proceedings of the Royal Society A} (online; forthcoming in print).

[46] S. Weinstein, ``Multiple time dimensions'', \url{arXiv:0812.3869v1 (2008)}.

\end{document}